\documentclass[%
 aip,
 amsmath,amssymb,
 reprint,%
]{revtex4-1}

\usepackage{graphicx}
\usepackage{dcolumn}
\usepackage{bm}

\usepackage[utf8]{inputenc}
\usepackage[T1]{fontenc}
\usepackage{mathptmx}
\usepackage{xcolor}
\usepackage{hyperref}

\newcommand{\be}{\begin{equation}}
\newcommand{\ee}{\end{equation}}

\newcommand{\jm}{\mathrm{j}}
\newcommand{\nn}{\nonumber}

\begin{document}

\title{Tunable High-Q Resonator by General Impedance Converter}

\author{Toshiro~Mifune}
\author{Todor~M.~Mishonov}
\thanks{The authors to whom correspondence may be addressed:\\
mishonov@bgphysics.eu, varon@bgphysics.eu}
\author{Nikola~S.~Serafimov}
\author{Iglika~M.~Dimitrova}
\affiliation{Georgi Nadjakov Institute of Solid State Physics, Bulgarian Academy of Sciences,\\
72 Tzarigradsko Chaussee Blvd., BG-1784 Sofia, Bulgaria}
\author{Riste Popeski-Dimovski}
\author{Leonora Velkoska}
\affiliation{Institute of Physics, Faculty of Natural Sciences and Mathematics, 
Sts. Cyril and Methodius University,\\
3 Arhimedova Str., MKD-1000 Skopje, R.~N.~Macedonia}
\author{Emil~G.~Petkov}
\author{Albert~M.~Varonov}
\thanks{The authors to whom correspondence may be addressed:\\
mishonov@bgphysics.eu, varon@bgphysics.eu}
\author{Alberto~Barone}
\affiliation{Georgi Nadjakov Institute of Solid State Physics, Bulgarian Academy of Sciences,\\
72 Tzarigradsko Chaussee Blvd., BG-1784 Sofia, Bulgaria}

\date{1 June 2021, 17:20}

\begin{abstract}
For the need of measurements focused in condensed matter physics 
and especially Bernoulli effect in superconductors 
we have developed an active resonator with dual operational amplifiers.
A tunable high-Q resonator is performed in the schematics of the 
the General Impedance Converter (GIC).
In the framework of frequency dependent open-loop gain of operational amplifiers,
a general formula of the frequency dependence of the impedance of GIC is derived.
The explicit formulas for the resonance frequency and Q-factor include
as immanent parameter the crossover frequency of the operational amplifier.
Voltage measurements of GIC with a lock-in 
amplifier perfectly agree with the derived formulas.
A table reveals that electrometer operational amplifiers are the best choice 
to build the described resonator.
\end{abstract}

\maketitle

\begin{figure}[h]%
\centering
\includegraphics[scale=0.35]{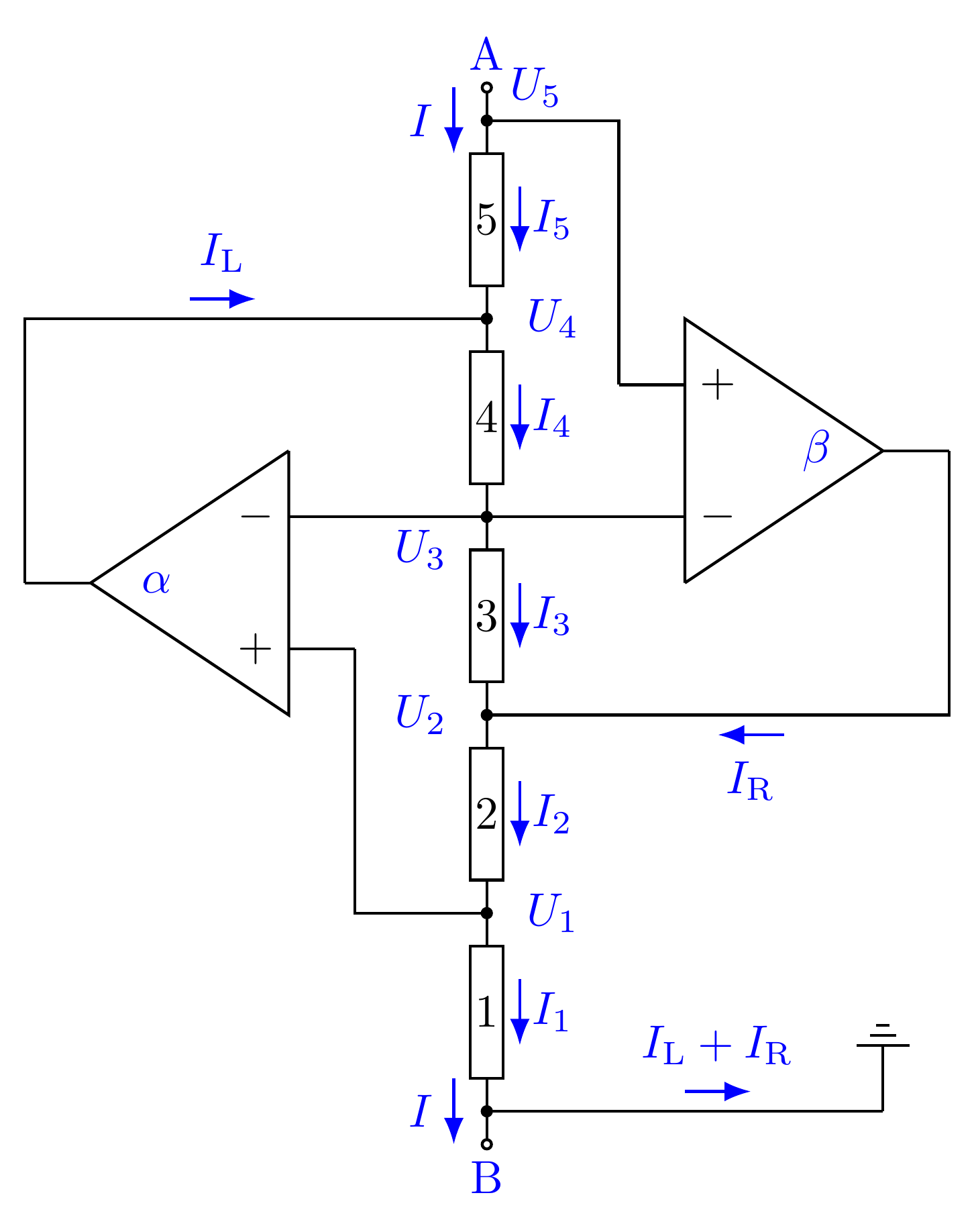}
\caption{
General Impedance Converter (GIC);
the impedance $Z=U/I$ is between points A and B.
$U=U_\mathrm{A}-U_\mathrm{B}$ is the voltage difference between those
electrodes and $I$ is the corresponding current.
The common point of the GIC (drawn intentionally upside down)
is not connected with the common point of the rest 
of the circuit and in this sense we consider floating connection of GIC
(see PCB drawing in supplementary material).
}
\label{Fig:GIC}
\end{figure}

High-Q resonators with resonance frequency $f_\mathrm{res}$
can find many technical applications
for which it is necessary to study the frequency dependence of a signal 
and simultaneously it is necessary this signal to be filtered by this high-Q resonator.
The purpose of the present study is to represent a new possible solution of this problem
in which the resonator is performed by 2 operational amplifiers (OpAmp)
included in the well-known topology of the 
General Impedance Converter (GIC)~\cite{Riordan:67,Antoniou:69,Sun,Franco,Schaumann:01,AD712}
drawn in Fig.~\ref{Fig:GIC}.
The work of the device is well described
by the single pole approximation of the open-loop gain of a operational amplifier
\begin{align}
G(\omega)\approx f_c/\jm f,
\quad\mbox{for} \quad f_c/G_0\ll f \ll f_c,
\end{align}
where $f_c$ is the crossover frequency, $G_0\sim 10^5$ is the static open-loop gain,
$f \equiv \omega/2\pi$ is the frequency, $\omega$ is the angular frequency,
and $\jm$ is the imaginary unit.
Let us recall also the common relation between
the plus $U_+$ and minus $U_-$ voltages of an OpAmp and the output one $U_\mathrm{o}$
\begin{align}
\label{master}
\alpha U_\mathrm{o}=U_+ - U_-,\quad
\alpha(\omega)  \equiv 1/G(\omega) \approx \alpha_0+\tau s+\gamma s^2,
\end{align}
where $s\equiv\mathrm{j}\omega$ is a widely used notation in electronics,
$\alpha_0 \equiv 1/G_0$
and the time constant 
$\tau=1/\omega_c\equiv1/2\pi f_c$ is a convenient parametrization of the crossover frequency~\cite{Ragazzini,master,Ghausi} and Ref.~\onlinecite[Eq.~(6.3)]{Franco}.
The linear dependence of the reciprocal open loop gain $\alpha\approx\tau s$
is often used in many specifications of OpAmp,
see for example Ref.~\onlinecite{ADA4817} and cited there
frequency dependent formulas for the amplification of inverting
$A_\mathrm{inv}(\omega)=-1/[(r/R+1)\,\alpha+r/R]$
and non-inverting 
$A_\mathrm{non}(\omega)=1/[\alpha+1/(R/r+1)]$
amplifiers,
where $R$ is the feedback resistance and $r$ is the gain resistance.


The schematics of the GIC analyzed in the present paper is shown in Fig.~\ref{Fig:GIC},
where 5 impedances $Z_1,\,Z_2,\dots,\, Z_5$, voltages $U_0,\,U_1,\,\dots,\, U_5$,
and currents $I_1,\,I_2,\dots,\, I_5$ are represented.
For convenience $U_5\equiv U_\mathrm{A}$ and $U_0\equiv U_\mathrm{B}=0.$ 
The current through the impedances and voltages are related by the Ohm law
\begin{align}
&
U_1-U_0=Z_1I_1,\quad
U_2-U_1=Z_2I_2,\\
&
U_3-U_2=Z_3I_3,\quad
U_4-U_3=Z_4I_4,\quad
U_5-U_4=Z_5I_5. \nn
\end{align}
We consider the input currents at the voltage inputs of the OpAmps 
negligible, which gives
$I_2=I_1$, $I_4=I_3$ and $I_5=I$.
The master equation Eq.~(\ref{master}) applied to both OpAmps
gives the last equations of the system
\begin{equation}
U_1-U_3=\alpha U_4,\qquad
U_5-U_3=\beta U_2.
\end{equation}
We suppose the use of a double OpAmp for which $\alpha\approx\beta$.
Taking into account also the re-notation $U=U_5=U_\mathrm{A}$ 
and $U_0=U_\mathrm{B}=0,$
the solution of the simple system of equations 
\begin{align}
\label{Z(omega)}
Z(\omega) \equiv \dfrac{U}{I}
&=\dfrac{Z_5}
{1-\dfrac{(Z_1+Z_2)-(Z_3+Z_4)K}{(1+\beta)(Z_1+Z_2)-Z_3K}}, \nn \\
K & \equiv\frac{Z_2+ (Z_1+Z_2)\alpha}{Z_3+(Z_3+Z_4)\alpha}.
\end{align}
At low frequencies $f\ll f_c$ and negligible $\alpha_0$,
i.e. in the infinite open-loop gain approximation $U_+\approx U_-$
this general formula gives the well-known
low frequency approximation
$Z(\omega\rightarrow 0)\approx Z_1Z_3Z_5/Z_2Z_4.$

In the present work we analyze the case when 4 of the impedances of a GIC are resistors
$Z_1=r_1,$ $Z_2=r_2,$ $Z_3=r_3,$  $Z_5=r_5$ 
and only one of them is a capacitor $Z_4=1/\mathrm{j}\omega C_4$;
metalized plastic thin films (polyester or polypropylene) which has dielectric loses of order $10^{-4}$.
At low frequencies this approximation gives
$Z=\mathrm{j}\omega L,$ where $L=C_4r_1r_3r_5/r_2.$
Actually simulated inductances together with D-elements
\cite{Franco,Zumbahlen@GIC,Lutz1,Lutz2} are the main
applications of GIC.
In our example $r_5=r_3=1\,\mathrm{k}\Omega,$
$r_2=r_1=100\,\Omega,$ and 
metallized polyester film capacitor
$C_4=10\;\mu\mathrm{F},$
which gives $L=10\,$H.
This set-up was given at the 7
Experimental Physics Olympiad.~\cite{epo7:a}

\begin{figure}[h]%
\centering
\includegraphics[scale=0.35]{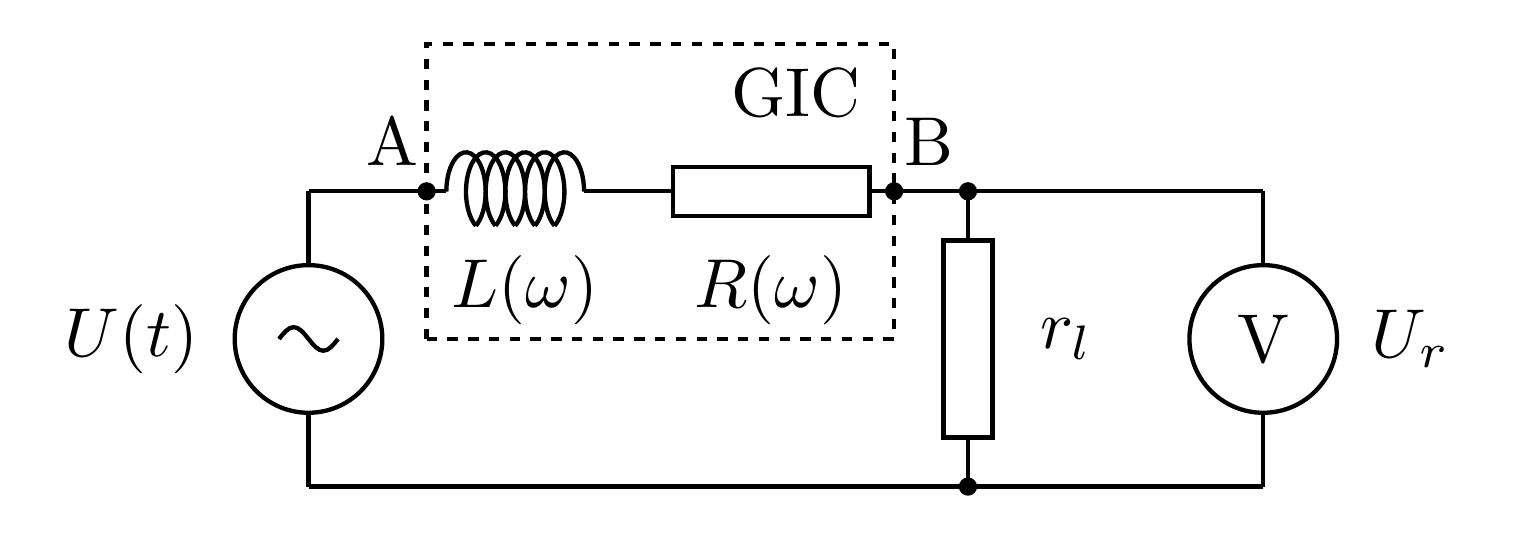}
\caption{Circuit for $Z(\omega)$ measurement with Anfatec USB~Lock--in~250~kHz~amplifier.~\cite{LockIn}
A sine voltage $U(t)$ from the lock-in amplifier is applied to GIC with impedance
$Z(\omega) = R(\omega)+\jm \omega L(\omega)$ and a serially connected load resistor~$r_l$.
The voltage drop $U_r$ on $r_l$ is measured 
by the lock-in for a range of frequencies.}
\label{Fig:Lock-In}
\end{figure}
\begin{figure}[h]%
\centering
\includegraphics[scale=0.4]{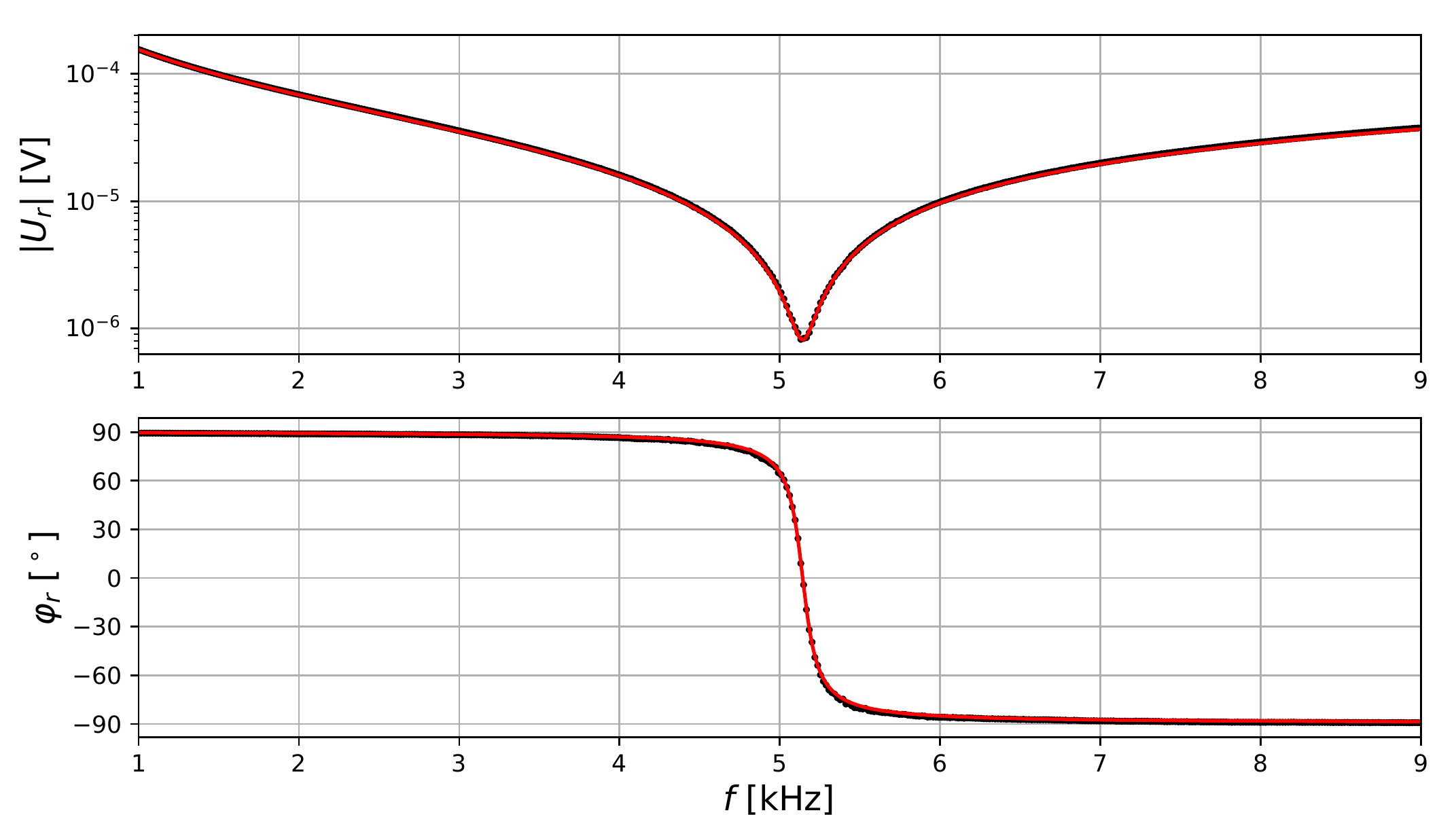}
\caption{
Voltage on the load resistor $r_l$ as a function of the frequency $f$. 
Points are experimental data,
lines are calculated by our analytical results
Eq.~(\ref{Z(omega)}) and Eq.~(\ref{U(omega)}).
Up: frequency dependence of the modulus of the voltage 
in log scale $|U_r(f)|$.
The deep cusp corresponds to a high-Q resonance.
Down: frequency dependence of the phase $\varphi_r(f)$.
In a narrow frequency region $\sim f_\mathrm{res}/\mathcal{Q}$,
close to $f_\mathrm{res}$, the phase decreases by 180$^\circ$.
}
\label{Fig:U_Phi}
\end{figure}

If the GIC is sequentially connected to a load resistor $r_l$,
as presented in Fig.~\ref{Fig:Lock-In},
the sequential impedance becomes $Z_s\equiv Z+r_l$.
For applied harmonic voltage $U(t)$
the current $I=U/Z_s$ 
and the voltage on the load resistor
\begin{align}
& U_r =\dfrac{U}{Z_s} r_l 
=|U_r|\mathrm{e}^{\mathrm{j}\varphi_r}
=U_r^\prime+\jm U_r^{\prime\prime}, \\
& \varphi_r(\omega) =\arctan\left(\frac{U_r^{\prime\prime}}{U_r^\prime}\right)
\in\left(-\dfrac{\pi}{2},\,\dfrac{\pi}{2}\right), \\
& |U_r(\omega)|=\sqrt{(U_r^\prime)^2+(U_r^{\prime\prime})^2}.
\label{U(omega)}
\end{align} 
Using a USB lock-in amplifier\cite{LockIn}
with $U=1\,$V and $r_l=10\,\Omega$ we measure the frequency dependence of the modulus 
$|U_r|$ and the phase $\varphi_r$ with the experimental set-up in Fig.~\ref{Fig:Lock-In}.
A sine voltage $U(t)$ is applied to the inductance with impedance
$Z(\omega) = R(\omega)+\jm \omega L(\omega)$ 
and the serially connected to it resistor $r_l$.
The voltage drop $U_r$ on $r_l$ is measured by the lock-in amplifier for different frequencies.
The experimental data and the fit according to our analytical results
Eq.~(\ref{Z(omega)}) and Eq.~(\ref{U(omega)}) are shown in Fig.~\ref{Fig:U_Phi}.
The fit of our theoretical formulas
Eq.~(\ref{Z(omega)}) and Eq.~(\ref{U(omega)})
to the experimental data for the
frequency dependence modulus $|U_r|$ and phase $\varphi_r$ 
gives as fitting parameters of the used AD712KN~\cite{AD712}
$f_c=3.3$~MHz and $G_0 \approx 26 \times 10^3$.

One can see a very narrow resonance with high Q-factor 
$\mathcal{Q} \sim 10^2$ at a resonance frequency $f_\mathrm{res}=5.1\,$kHz.
Having such an excellent fit of the experimental voltage with the analytical formula 
Eq.~(\ref{Z(omega)}),
one can easily calculate the real and imaginary part of the impedance of the GIC and to compare with the processed data for the impedance
\begin{align}
\label{complex}
\frac{Z(\omega)}{r_l}&=\frac{Z^{\prime}+\mathrm{j}Z^{\prime\prime}}{r_l}
=\frac{|Z|\mathrm{e}^{\mathrm{j}\varphi}}{r_l}
=\left(\dfrac{U}{U_r}-1\right) \\
& =\left(\dfrac{U}{|U_r|}\mathrm{e}^{-\mathrm{j}\varphi_r}\!-\!1\right)
=-A_\mathrm{inv}(0)=\left(A_\mathrm{non}(0)\!-\!1\right),\nn
\\
\label{real_imaginary}
Z^{\prime}(\omega)&=R(\omega)=\left(\dfrac{U}{|U_r|}\cos(\varphi_r)-1\right)r_l, \\
 Z^{\prime\prime}(\omega)&=-r_l\dfrac{U}{|U_r|}\sin(\varphi_r)
=\omega L(\omega)
=-\dfrac{1}{\omega C(\omega)}, \nn \\
\label{L_vs_C}
&L(\omega<\omega_\mathrm{res})>0,\qquad
C(\omega>\omega_\mathrm{res})>0. \nn
\end{align}
In such a way we have expressed the impedance $Z(\omega)$ 
of GIC by the measurable variables
of modulus $|U_r|$ and phase $\varphi_r$
measured by a lock-in on a load resistor $r_l$
sequentially connected to GIC.
Below $f_\mathrm{res}$ GIC is an inductance, 
while for frequencies higher than $f_\mathrm{res}$ GIC has
frequency dependence as a capacitor. 
Qualitatively we have a parallel resonance circuit 
with a capacitor and inductance.
It is remarkable that such a Q-factor can be easily reached even at 
audio frequencies.
The phase changes at 180$^\circ$ in frequency interval of only 100~Hz.

For $\omega_\mathrm{res} \equiv 2\pi f_\mathrm{res}$ and
$\mathcal{Q}$ it is possible to obtain 
explicit analytical expressions by calculating the condition of zero conductivity 
$1/Z(\omega_\mathrm{res})=0$.
Introducing 
\begin{align}
& \Omega\equiv\omega\tau,
\quad S=\mathrm{j} \Omega= \mathrm{j} \frac{f}{f_c}=s\tau,\quad
\rho_4\equiv\dfrac{\tau}{C_4}\ll r_3, \nn \\
& \epsilon_4\equiv \frac{\rho_4}{r_3}=\frac{\tau}{r_3C_4}\ll1, \quad
K_1 \equiv \frac{r_1}{r_1+r_2}, \quad
K_2 \equiv \frac{r_2}{r_1+r_2} 
\nn
\end{align}
after some algebra from Eq.~(\ref{Z(omega)}) the impedance
\begin{widetext}
\be
\label{Pade}
\frac{Z(\omega)}{r_5} =  \frac{S^3 + [(1+\epsilon_4)+2 a] S^2
+[a^2+a+K_1 +(1 + 2 a)\epsilon_4 ] S
+ (a^2+a)\epsilon_4 }
{S^3 + [(1+\epsilon_4)+2 a] S^2
+[a^2 + a +(1 + 2 a)\epsilon_4 ] S
+ (a^2 + a + K_2)\epsilon_4 },
\quad a(\omega)\equiv\alpha_0+\gamma S^2.
\ee
\end{widetext}
The explicit formulas for the frequency dependence of the impedance of GIC 
Eq.~(\ref{Pade}) and Eq.~(\ref{Z(omega)}) give the answer to every 
question related to the linear theory of GIC.
The zeros of the denominator of Eq.~(\ref{Pade}) describe the resonances of the impedance,
where the influence of $\gamma \ll 1$ is negligible
and the annulation of the denominator
using only the linear terms of $a$ and $\varepsilon_2$ gives
\be
\label{eigen-frequency}
S^3+
(1+2a+\epsilon_4)S^2+
(a+\epsilon_4) S+
\epsilon_4 K_2 = 0.
\ee
This equation has an approximate solution at 
$\Omega\approx (1+\jm/2\mathcal{Q})\,\Omega_\mathrm{res}, \quad
\Omega_\mathrm{res} \approx \sqrt{\epsilon_4 K_2}
=\tau\omega_\mathrm{res}\ll1$
and
\begin{align}
\label{res_frequency}
& f_\mathrm{res}=\frac{\omega_\mathrm{res}}{2\pi}\approx
\dfrac{1}{\sqrt{2\pi(1+r_1/r_2)}}\sqrt{\frac{f_c}{r_3C_4}}
\ll f_c, \\
\label{Q-factor}
& \mathcal{Q}\approx
\frac{\sqrt{K_2\varepsilon_4}}{\alpha_0+K_1\varepsilon_4}
\gg 1, 
\end{align}
i.e. in order to increase Q-factor is necessary to use simultaneously 
high open loop gain $G_0=1/\alpha_0$ and high crossover frequency $f_c$.
In some sense these results are derived by the
Manhattan equation\cite{Manhattan} in action.

For optimized resonator $(\omega_c/G_0)r_3C_4\sim K_1.$
The expression for $f_\mathrm{res}$ reveals that $r_3$ is the most convenient tunable parameter, while $r_1$ can be used for fine tuning.
For $r_1\ll r_2$ when $K_2\approx 1$ we have 
$\Omega_\mathrm{res}=\sqrt{\epsilon_4}\ll1,$
i.e some time dependent variable $X(t)$ obeys the oscillator equation
$\mathrm{d}^2 X(t)/\mathrm{d}t^2=-X(t)/r_3C_4\tau$
and the problem deserves more ingenious analysis
giving the simple result $\omega_\mathrm{res}=1/\sqrt{\tau r_3C_4}$
directly.

For very low frequencies Eq.~(\ref{Pade}) gives Ohmic resistance
\be
R_L\equiv Z^\prime(\omega\rightarrow 0)=(1+r_1/r_2)r_5/G_0\ll r_5,
\ee
an we have practically an ideal inductor with $L=10$~H.
We wish to emphasize that our results in Eqs.~(\ref{res_frequency}) and (\ref{Q-factor})
are based on the frequency dependent open-loop gain  for the OpAmp described by 
Eq.~(\ref{master}).

%
The result for the Q-factor Eq.~(\ref{Q-factor}) deserves a detailed analysis.
According to the Manhattan equation\cite{Manhattan} for the open-loop gain 
Eq.~(\ref{master}) $G(f) \approx (1/G_0+\jm f/f_c)^{-1}$ 
the crossover frequency\cite{ADA4817} is defined 
as a frequency for which $|G(f_c)|=1$, 
that is why this frequency is also called unity gain frequency.\cite{Dostal}
In the same textbook by Dostal~\onlinecite[Chap.~2]{Dostal} 
is introduced the dominant frequency 
\be
f_d\equiv f_c/G_0
\ee 
according to which the open-loop gain in power 
decreases twice with respect to the zero frequency
$|G(f_d)|^2=G_0^2/2.$
Using these notions Eq.~(\ref{Q-factor}) reads
\be
\mathcal{Q}=\dfrac{\dfrac{f_\mathrm{res}}{f_d}}
{1+\dfrac{r_1}{r_2} 
\left(\dfrac{f_\mathrm{res}}{f_c}\right)\left(\dfrac{f_\mathrm{res}}{f_d}\right)}\gg1.
\label{Q_fres}
\ee
For special case of $r_1=r_2$ introducing $x\equiv f_\mathrm{res}/f_c$
the Q-factor writes $\mathcal{Q}(x)=x/(G_0^{-1}+x^2)$. 
This function has a maximum 
$\mathcal{Q}_\mathrm{max}=\sqrt{G_0}/2$
at $x_\mathrm{max}=1/\sqrt{G_0}$, i.e. at
$f_\mathrm{max}=f_c/\sqrt{G_0}=\sqrt{f_df_c}$ is the
resonance frequency at $\mathcal{Q}_\mathrm{max}$.
Usually the unity gain frequency $f_c$ is so high that the second term in the denominator gives only several percent contribution and with acceptable for the choice of OpAmps,
we can use the approximation
\be
\mathcal{Q}\approx f_\mathrm{res}/f_d\gg1.
\label{Q_qualitative}
\ee
In other words the dominant frequency 
$f_d=f_c/G_0$ is the most important parameter 
for the choice of OpAmp used for the described resonator.
For an ideal OpAmp with $f_c=\infty$ and $G_0=\infty$
this resonance cannot be described;
it is necessary to precise also that $f_d=f_c/G_0=0$
which gives ideal $\mathcal{Q}=\infty$.
In Table~\ref{tbl:fd} the parameters for the frequently used OpAmps are given.
\begin{table}[h]
\begin{tabular}{ l  r r  r  r}
		\tableline \tableline
		&  \\ [-1em]
		OpAmp & \hspace{1.0pt} $f_c$~[MHz] & \hspace{1.0pt} $G_0$~\{$10^6$\} &   \hspace{1.0pt} $f_d$~[Hz] \hspace{2.5pt} & Reference \\ \tableline 
			&  \\ [-1em] 
			ADA4530 &   2   &   14 & 0.14 & ~~\onlinecite[Table~1, Fig.~55]{ADA4530} \\
			AD549     &   1   &  1.0 &   1.0 & ~~\onlinecite[Fig~8, Fig.~14]{AD549} \\
			AD8544	&	1   &  0.5 &   2.0 & ~~\onlinecite[Table~1, Fig.~18]{AD8544}  \\			
			TL072	     &	5	&  1.0 & 5.0 & ~~\onlinecite[p.~15, Fig.~6-8]{TL072} \\
			AD712	&	4	&  0.4 & 10 & ~~\onlinecite[Table~1, Fig.~11]{AD712} \\
			ADA4610 & 9.3& 0.1 & 93& ~~\onlinecite[Table~2, Fig.~26]{ADA4610} \\
			LTC6269	&	300&  0.25 & 1~200 & ~~\onlinecite[Table~2, G21]{LTC6269} \\
			ADA4898 & 100 & 0.05& 2~000 & ~~\onlinecite[Table~1, Fig.~19]{ADA4898} \\
			ADA4817 & 400 &0.0014& 286$\times 10^3$ & ~~\onlinecite[Table~2, Fig.~31]{ADA4817} \\
\tableline \tableline
\end{tabular}
\caption{According to the approximate Eq.~(\ref{Q_qualitative})} the best Q-factor is reached by the lowest dominant frequencies $f_d$. 
The table reveals that electrometer OpAmps are the best choice to build the described resonators.
On the other hand low-noise and high frequency amplifiers are unsuitable for this purpose.
The estimation according to Eq.~\ref{Q_fres} and further analysis show that
one can expect
with ADA4530\cite{ADA4530} one can reach $\mathcal{Q}>1000$ and
with the other electrometer AD549\cite{AD549} $\mathcal{Q} \approx 500$ both at kHz range.
	\label{tbl:fd}
\end{table}

%


Usually finite frequency of the crossover frequency $f_c$ is considered as some 
non-ideality of an OpAmp but the purpose of the present paper 
is to demonstrate that this inequality can be used for something useful,
to design and create tunable high-Q resonators for various applications.\cite{Munoz:05,Munoz:06,Montero:07,Ruse:04}
In other words, the novelty of the proposed scheme is based on the immanent property of the operational amplifiers -- the finite crossover frequency $f_c$.
Recently we have also developed a method for fast and accurate measurement of the crossover frequency of operational amplifiers.\cite{PDF}
All these studies are part of a creation of instrument for measurements focused in condensed matter physics and especially Bernoulli effect in superconductors.\cite{Varna}
In conclusion in order to reach $\mathcal{Q}>100$ it is recommended to use 
a contemporary electrometer OpAmp.


Toshiro Mifune and Alberto Barone 
wish to thank to Hassan Chamati for the
hospitality in the Georgi Nadjakov Institute of Solid State Physics and creative atmosphere during this long time endeavor.
Albert Varonov would like to acknowledge 
the support by the Joint Institute for Nuclear Research, Dubna, Russian
Federation - THEME 01-3-1137-2019/2023 and Grant No D01-378/18.12.2020 of
the Ministry of Education and Science of Bulgaria.



\bibliography{GIC}

\begin{thebibliography}{30}%
\makeatletter
\providecommand \@ifxundefined [1]{%
 \@ifx{#1\undefined}
}%
\providecommand \@ifnum [1]{%
 \ifnum #1\expandafter \@firstoftwo
 \else \expandafter \@secondoftwo
 \fi
}%
\providecommand \@ifx [1]{%
 \ifx #1\expandafter \@firstoftwo
 \else \expandafter \@secondoftwo
 \fi
}%
\providecommand \natexlab [1]{#1}%
\providecommand \enquote  [1]{``#1''}%
\providecommand \bibnamefont  [1]{#1}%
\providecommand \bibfnamefont [1]{#1}%
\providecommand \citenamefont [1]{#1}%
\providecommand \href@noop [0]{\@secondoftwo}%
\providecommand \href [0]{\begingroup \@sanitize@url \@href}%
\providecommand \@href[1]{\@@startlink{#1}\@@href}%
\providecommand \@@href[1]{\endgroup#1\@@endlink}%
\providecommand \@sanitize@url [0]{\catcode `\\12\catcode `\$12\catcode
  `\&12\catcode `\#12\catcode `\^12\catcode `\_12\catcode `\%12\relax}%
\providecommand \@@startlink[1]{}%
\providecommand \@@endlink[0]{}%
\providecommand \url  [0]{\begingroup\@sanitize@url \@url }%
\providecommand \@url [1]{\endgroup\@href {#1}{\urlprefix }}%
\providecommand \urlprefix  [0]{URL }%
\providecommand \Eprint [0]{\href }%
\providecommand \doibase [0]{http://dx.doi.org/}%
\providecommand \selectlanguage [0]{\@gobble}%
\providecommand \bibinfo  [0]{\@secondoftwo}%
\providecommand \bibfield  [0]{\@secondoftwo}%
\providecommand \translation [1]{[#1]}%
\providecommand \BibitemOpen [0]{}%
\providecommand \bibitemStop [0]{}%
\providecommand \bibitemNoStop [0]{.\EOS\space}%
\providecommand \EOS [0]{\spacefactor3000\relax}%
\providecommand \BibitemShut  [1]{\csname bibitem#1\endcsname}%
\let\auto@bib@innerbib\@empty
\bibitem [{\citenamefont {Riordan}(1967)}]{Riordan:67}%
  \BibitemOpen
  \bibfield  {author} {\bibinfo {author} {\bibfnamefont {R.}~\bibnamefont
  {Riordan}},\ }\bibfield  {title} {\enquote {\bibinfo {title} {Simulated
  inductors using differential amplifiers},}\ }\href {\doibase
  10.1049/el:19670039} {\bibfield  {journal} {\bibinfo  {journal} {El. Lett.}\
  }\textbf {\bibinfo {volume} {3}},\ \bibinfo {pages} {50--51} (\bibinfo {year}
  {1967})}\BibitemShut {NoStop}%
\bibitem [{\citenamefont {{Antoniou}}(1969)}]{Antoniou:69}%
  \BibitemOpen
  \bibfield  {author} {\bibinfo {author} {\bibfnamefont {A.}~\bibnamefont
  {{Antoniou}}},\ }\bibfield  {title} {\enquote {\bibinfo {title} {Realisation
  of gyrators using operational amplifiers, and their use in rc-active-network
  synthesis},}\ }\href {\doibase 10.1049/piee.1969.0339} {\bibfield  {journal}
  {\bibinfo  {journal} {Proc. IEE}\ }\textbf {\bibinfo {volume} {116}},\
  \bibinfo {pages} {1838--1850} (\bibinfo {year} {1969})}\BibitemShut {NoStop}%
\bibitem [{\citenamefont {T.~Deliyannis}\ and\ \citenamefont
  {Fiddler}(1999)}]{Sun}%
  \BibitemOpen
  \bibfield  {author} {\bibinfo {author} {\bibfnamefont {Y.~S.}\ \bibnamefont
  {T.~Deliyannis}}\ and\ \bibinfo {author} {\bibfnamefont {J.~K.}\ \bibnamefont
  {Fiddler}},\ }\href@noop {} {\emph {\bibinfo {title} {Continuous-Time Active
  Filter Design}}}\ (\bibinfo  {publisher} {CRC-Press},\ \bibinfo {address}
  {New York},\ \bibinfo {year} {1999})\BibitemShut {NoStop}%
\bibitem [{\citenamefont {Franco}(2002)}]{Franco}%
  \BibitemOpen
  \bibfield  {author} {\bibinfo {author} {\bibfnamefont {S.}~\bibnamefont
  {Franco}},\ }\enquote {\bibinfo {title} {Design with operational amplifiers
  and analog integrated circuits},}\ \ (\bibinfo  {publisher} {McGraw Hill},\
  \bibinfo {address} {New York},\ \bibinfo {year} {2002})\ Chap.~\bibinfo
  {chapter} {4},\ \bibinfo {edition} {3rd}\ ed.\BibitemShut {Stop}%
\bibitem [{\citenamefont {Schaumann}\ and\ \citenamefont
  {Valkenburg}(2001)}]{Schaumann:01}%
  \BibitemOpen
  \bibfield  {author} {\bibinfo {author} {\bibfnamefont {R.}~\bibnamefont
  {Schaumann}}\ and\ \bibinfo {author} {\bibfnamefont {M.~E.~V.}\ \bibnamefont
  {Valkenburg}},\ }\enquote {\bibinfo {title} {Design of analog filters},}\ \
  (\bibinfo  {publisher} {Oxford University Press},\ \bibinfo {address} {New
  York},\ \bibinfo {year} {2001})\ Chap.~\bibinfo {chapter} {4}\BibitemShut
  {NoStop}%
\bibitem [{\citenamefont {Anonym}(2018)}]{AD712}%
  \BibitemOpen
  \bibfield  {author} {\bibinfo {author} {\bibnamefont {Anonym}},\ }\href
  {https://www.analog.com/media/en/technical-documentation/data-sheets/AD712.pdf}
  {\enquote {\bibinfo {title} {Precision, low cost, high speed {BiFET} dual op
  amp {AD712}},}\ }\bibinfo {type} {datasheet}\ \bibinfo {number} {Rev. I}\
  (\bibinfo  {institution} {Analog Devices Inc.},\ \bibinfo {year}
  {2018})\BibitemShut {NoStop}%
\bibitem [{\citenamefont {{Ragazzini}}, \citenamefont {{Randall}},\ and\
  \citenamefont {{Russell}}(1947)}]{Ragazzini}%
  \BibitemOpen
  \bibfield  {author} {\bibinfo {author} {\bibfnamefont {J.~R.}\ \bibnamefont
  {{Ragazzini}}}, \bibinfo {author} {\bibfnamefont {R.~H.}\ \bibnamefont
  {{Randall}}}, \ and\ \bibinfo {author} {\bibfnamefont {F.~A.}\ \bibnamefont
  {{Russell}}},\ }\bibfield  {title} {\enquote {\bibinfo {title} {Analysis of
  problems in dynamics by electronic circuits},}\ }\href {\doibase
  10.1109/JRPROC.1947.232616} {\bibfield  {journal} {\bibinfo  {journal} {Proc.
  IRE}\ }\textbf {\bibinfo {volume} {35}},\ \bibinfo {pages} {444--452}
  (\bibinfo {year} {1947})}\BibitemShut {NoStop}%
\bibitem [{\citenamefont {Mishonov}\ \emph
  {et~al.}(2019{\natexlab{a}})\citenamefont {Mishonov}, \citenamefont
  {Danchev}, \citenamefont {Petkov}, \citenamefont {Gourev}, \citenamefont
  {Dimitrova}, \citenamefont {Serafimov}, \citenamefont {Stefanov},\ and\
  \citenamefont {Varonov}}]{master}%
  \BibitemOpen
  \bibfield  {author} {\bibinfo {author} {\bibfnamefont {T.~M.}\ \bibnamefont
  {Mishonov}}, \bibinfo {author} {\bibfnamefont {V.~I.}\ \bibnamefont
  {Danchev}}, \bibinfo {author} {\bibfnamefont {E.~G.}\ \bibnamefont {Petkov}},
  \bibinfo {author} {\bibfnamefont {V.~N.}\ \bibnamefont {Gourev}}, \bibinfo
  {author} {\bibfnamefont {I.~M.}\ \bibnamefont {Dimitrova}}, \bibinfo {author}
  {\bibfnamefont {N.~S.}\ \bibnamefont {Serafimov}}, \bibinfo {author}
  {\bibfnamefont {A.~A.}\ \bibnamefont {Stefanov}}, \ and\ \bibinfo {author}
  {\bibfnamefont {A.~M.}\ \bibnamefont {Varonov}},\ }\bibfield  {title}
  {\enquote {\bibinfo {title} {Master equation for operational amplifiers:
  stability of negative differential converters, crossover frequency and
  pass-bandwidth},}\ }\href {\doibase 10.1088/2399-6528/ab050b} {\bibfield
  {journal} {\bibinfo  {journal} {J. Phys. Comm.}\ }\textbf {\bibinfo {volume}
  {3}},\ \bibinfo {pages} {035004} (\bibinfo {year}
  {2019}{\natexlab{a}})}\BibitemShut {NoStop}%
\bibitem [{\citenamefont {Ghausi}\ and\ \citenamefont {Laker}(1991)}]{Ghausi}%
  \BibitemOpen
  \bibfield  {author} {\bibinfo {author} {\bibfnamefont {M.~S.}\ \bibnamefont
  {Ghausi}}\ and\ \bibinfo {author} {\bibfnamefont {K.~R.}\ \bibnamefont
  {Laker}},\ }\href {\doibase 10.1049/SBCS005E} {\emph {\bibinfo {title}
  {Modern Filter Design: Active RC and Switched Capacitor}}}\ (\bibinfo
  {publisher} {Prentice-Hall},\ \bibinfo {address} {Englewood Cliffs, NJ},\
  \bibinfo {year} {1991})\BibitemShut {NoStop}%
\bibitem [{\citenamefont {Anonym}(2019{\natexlab{a}})}]{ADA4817}%
  \BibitemOpen
  \bibfield  {author} {\bibinfo {author} {\bibnamefont {Anonym}},\ }\href
  {https://www.analog.com/media/en/technical-documentation/data-sheets/ADA4817-1_4817-2.pdf}
  {\enquote {\bibinfo {title} {Low noise, 1 ghz, {FastFET} op amps
  {ADA4817-1/ADA4817-2}},}\ }\bibinfo {type} {datasheet}\ \bibinfo {number}
  {Rev. G}\ (\bibinfo  {institution} {Analog Devices Inc.},\ \bibinfo {year}
  {2019})\BibitemShut {NoStop}%
\bibitem [{\citenamefont {Inc.}(2008)}]{Zumbahlen@GIC}%
  \BibitemOpen
  \bibfield  {author} {\bibinfo {author} {\bibfnamefont {A.~D.}\ \bibnamefont
  {Inc.}},\ }\enquote {\bibinfo {title} {Linear circuit design handbook},}\ \
  (\bibinfo  {publisher} {Newnes/Elsevier},\ \bibinfo {address} {New York},\
  \bibinfo {year} {2008})\ Chap.~\bibinfo {chapter} {8}, pp.\ \bibinfo {pages}
  {8.68--8.69},\ \bibinfo {note} {{GIC} Figs.~8.45 and 8.46}\BibitemShut
  {NoStop}%
\bibitem [{\citenamefont {von Wangenheim}(1996)}]{Lutz1}%
  \BibitemOpen
  \bibfield  {author} {\bibinfo {author} {\bibfnamefont {L.}~\bibnamefont {von
  Wangenheim}},\ }\bibfield  {title} {\enquote {\bibinfo {title} {{Modification
  of the classical GIC structure and its application to RC-oscillators}},}\
  }\href
  {https://digital-library.theiet.org/content/journals/10.1049/el_19960041}
  {\bibfield  {journal} {\bibinfo  {journal} {Electronics Letters}\ }\textbf
  {\bibinfo {volume} {32}},\ \bibinfo {pages} {6--8(2)} (\bibinfo {year}
  {1996})}\BibitemShut {NoStop}%
\bibitem [{\citenamefont {Von~Wangenheim}(1997)}]{Lutz2}%
  \BibitemOpen
  \bibfield  {author} {\bibinfo {author} {\bibfnamefont {L.}~\bibnamefont
  {Von~Wangenheim}},\ }\bibfield  {title} {\enquote {\bibinfo {title} {Pspice
  tunes oscillator circuits},}\ }\href@noop {} {\bibfield  {journal} {\bibinfo
  {journal} {EDN}\ }\textbf {\bibinfo {volume} {42}},\ \bibinfo {pages}
  {121--122} (\bibinfo {year} {1997})}\BibitemShut {NoStop}%
\bibitem [{\citenamefont {Mishonov}\ \emph
  {et~al.}(2019{\natexlab{b}})\citenamefont {Mishonov}, \citenamefont
  {Popeski-Dimovski}, \citenamefont {Velkoska}, \citenamefont {Dimitrova},
  \citenamefont {Gourev}, \citenamefont {Petkov}, \citenamefont {Petkov},\ and\
  \citenamefont {Varonov}}]{epo7:a}%
  \BibitemOpen
  \bibfield  {author} {\bibinfo {author} {\bibfnamefont {T.~M.}\ \bibnamefont
  {Mishonov}}, \bibinfo {author} {\bibfnamefont {R.}~\bibnamefont
  {Popeski-Dimovski}}, \bibinfo {author} {\bibfnamefont {L.}~\bibnamefont
  {Velkoska}}, \bibinfo {author} {\bibfnamefont {I.~M.}\ \bibnamefont
  {Dimitrova}}, \bibinfo {author} {\bibfnamefont {V.~N.}\ \bibnamefont
  {Gourev}}, \bibinfo {author} {\bibfnamefont {A.~P.}\ \bibnamefont {Petkov}},
  \bibinfo {author} {\bibfnamefont {E.~G.}\ \bibnamefont {Petkov}}, \ and\
  \bibinfo {author} {\bibfnamefont {A.~M.}\ \bibnamefont {Varonov}},\
  }\href@noop {} {\enquote {\bibinfo {title} {{The Day of the Inductance.
  Problem of the 7-th Experimental Physics Olympiad, Skopje, 7 December
  2019}},}\ } (\bibinfo {year} {2019}{\natexlab{b}}),\ \Eprint
  {http://arxiv.org/abs/1912.07368} {arXiv:1912.07368 [physics.ed-ph]}
  \BibitemShut {NoStop}%
\bibitem [{\citenamefont {Anonym}(2013)}]{LockIn}%
  \BibitemOpen
  \bibfield  {author} {\bibinfo {author} {\bibnamefont {Anonym}},\ }\href
  {http://www.anfatec.net/downloads/USBLockIn/USBLockIn250_Manual.pdf}
  {\enquote {\bibinfo {title} {{USB} lockin 250, lockin amplifier, amplifier 10
  mhz to 250 khz},}\ }\bibinfo {type} {manual}\ \bibinfo {number} {Rev. 1.03}\
  (\bibinfo  {institution} {Anfatec Instruments AG},\ \bibinfo {year}
  {2013})\BibitemShut {NoStop}%
\bibitem [{\citenamefont {Mishonov}\ \emph
  {et~al.}(2019{\natexlab{c}})\citenamefont {Mishonov}, \citenamefont
  {Stefanov}, \citenamefont {Petkov}, \citenamefont {Dimitrova}, \citenamefont
  {Danchev}, \citenamefont {Gourev},\ and\ \citenamefont
  {Varonov}}]{Manhattan}%
  \BibitemOpen
  \bibfield  {author} {\bibinfo {author} {\bibfnamefont {T.~M.}\ \bibnamefont
  {Mishonov}}, \bibinfo {author} {\bibfnamefont {A.~A.}\ \bibnamefont
  {Stefanov}}, \bibinfo {author} {\bibfnamefont {E.~G.}\ \bibnamefont
  {Petkov}}, \bibinfo {author} {\bibfnamefont {I.~M.}\ \bibnamefont
  {Dimitrova}}, \bibinfo {author} {\bibfnamefont {V.~I.}\ \bibnamefont
  {Danchev}}, \bibinfo {author} {\bibfnamefont {V.~N.}\ \bibnamefont {Gourev}},
  \ and\ \bibinfo {author} {\bibfnamefont {A.~M.}\ \bibnamefont {Varonov}},\
  }\bibfield  {title} {\enquote {\bibinfo {title} {Manhattan equation for the
  operational amplifier},}\ }in\ \href {\doibase 10.1063/1.5091329} {\emph
  {\bibinfo {booktitle} {10th Jubilee International Conference of the Balkan
  Physical Union}}},\ Vol.\ \bibinfo {volume} {2075}\ (\bibinfo  {publisher}
  {AIP Publishing},\ \bibinfo {year} {2019})\ p.\ \bibinfo {pages}
  {160002}\BibitemShut {NoStop}%
\bibitem [{\citenamefont {Dost\'al}(1993)}]{Dostal}%
  \BibitemOpen
  \bibfield  {author} {\bibinfo {author} {\bibfnamefont {J.}~\bibnamefont
  {Dost\'al}},\ }\enquote {\bibinfo {title} {Operational amplifiers},}\ \
  (\bibinfo  {publisher} {Butterworth-Heinemann},\ \bibinfo {address} {New
  York},\ \bibinfo {year} {1993})\ Chap.\ \bibinfo {chapter} {2~Operational
  Amplifier Parameters},\ \bibinfo {edition} {2nd}\ ed.,\ \bibinfo {note}
  {sec.~2.1~Linear Parameters and Linear Model}\BibitemShut {NoStop}%
\bibitem [{\citenamefont {Anonym}(2017)}]{ADA4530}%
  \BibitemOpen
  \bibfield  {author} {\bibinfo {author} {\bibnamefont {Anonym}},\ }\href
  {https://www.analog.com/media/en/technical-documentation/data-sheets/ADA4530-1.pdf}
  {\enquote {\bibinfo {title} {Femtoampere input bias current electrometer
  amplifier {ADA4530-1}},}\ }\bibinfo {type} {datasheet}\ \bibinfo {number}
  {Rev. B}\ (\bibinfo  {institution} {Analog Devices Inc.},\ \bibinfo {year}
  {2017})\BibitemShut {NoStop}%
\bibitem [{\citenamefont {Anonym}(2015{\natexlab{a}})}]{AD549}%
  \BibitemOpen
  \bibfield  {author} {\bibinfo {author} {\bibnamefont {Anonym}},\ }\href
  {https://www.analog.com/media/en/technical-documentation/data-sheets/AD549.pdf}
  {\enquote {\bibinfo {title} {Ultralow input bias current operational
  amplifier {AD549}},}\ }\bibinfo {type} {datasheet}\ \bibinfo {number} {Rev.
  K}\ (\bibinfo  {institution} {Analog Devices Inc.},\ \bibinfo {year}
  {2015})\BibitemShut {NoStop}%
\bibitem [{\citenamefont {Anonym}(2011)}]{AD8544}%
  \BibitemOpen
  \bibfield  {author} {\bibinfo {author} {\bibnamefont {Anonym}},\ }\href
  {https://www.analog.com/media/en/technical-documentation/data-sheets/AD8541_8542_8544.pdf}
  {\enquote {\bibinfo {title} {{CMOS} rail-to-rail general-purpose amplifiers
  {AD8541/AD8542/AD8544}},}\ }\bibinfo {type} {datasheet}\ \bibinfo {number}
  {Rev. G}\ (\bibinfo  {institution} {Analog Devices Inc.},\ \bibinfo {year}
  {2011})\BibitemShut {NoStop}%
\bibitem [{\citenamefont {Anonym}(2020)}]{TL072}%
  \BibitemOpen
  \bibfield  {author} {\bibinfo {author} {\bibnamefont {Anonym}},\ }\href
  {https://www.ti.com/lit/ds/slos080p/slos080p.pdf?ts=1622454176068&ref_url=https%253A%252F%252Fwww.ti.com%252Fproduct%252FTL072}
  {\enquote {\bibinfo {title} {{TL07xx} low-noise {FET}-input operational
  amplifiers},}\ }\bibinfo {type} {datasheet}\ \bibinfo {number} {Rev. P}\
  (\bibinfo  {institution} {Texas Instruments Inc.},\ \bibinfo {year}
  {2020})\BibitemShut {NoStop}%
\bibitem [{\citenamefont {Anonym}(2019{\natexlab{b}})}]{ADA4610}%
  \BibitemOpen
  \bibfield  {author} {\bibinfo {author} {\bibnamefont {Anonym}},\ }\href
  {https://www.analog.com/media/en/technical-documentation/data-sheets/ADA4610-1_4610-2_4610-4.pdf}
  {\enquote {\bibinfo {title} {Low noise, precision, rail-to-rail output,
  {JFET} single/dual/quad op amps {ADA4610-1/ADA4610-2/ADA4610-4}},}\ }\bibinfo
  {type} {datasheet}\ \bibinfo {number} {Rev. I}\ (\bibinfo  {institution}
  {Analog Devices Inc.},\ \bibinfo {year} {2019})\BibitemShut {NoStop}%
\bibitem [{\citenamefont {Anonym}(2014)}]{LTC6269}%
  \BibitemOpen
  \bibfield  {author} {\bibinfo {author} {\bibnamefont {Anonym}},\ }\href
  {https://www.analog.com/media/en/technical-documentation/data-sheets/62689f.pdf}
  {\enquote {\bibinfo {title} {{LTC6268/LTC6269} ultra-low bias current {FET}
  input op amp},}\ }\bibinfo {type} {datasheet}\ (\bibinfo  {institution}
  {Linear Technology Corporation},\ \bibinfo {year} {2014})\BibitemShut
  {NoStop}%
\bibitem [{\citenamefont {Anonym}(2015{\natexlab{b}})}]{ADA4898}%
  \BibitemOpen
  \bibfield  {author} {\bibinfo {author} {\bibnamefont {Anonym}},\ }\href
  {http://www.analog.com/static/imported-files/data_sheets/ADA4898-1_4898-2.pdf}
  {\enquote {\bibinfo {title} {High voltage, low noise,low distortion,
  unity-gain stable, high speed op amp {ADA4898-1/ADA4898-2}},}\ }\bibinfo
  {type} {datasheet}\ \bibinfo {number} {Rev. E}\ (\bibinfo  {institution}
  {Analog Devices Inc.},\ \bibinfo {year} {2015})\BibitemShut {NoStop}%
\bibitem [{\citenamefont {Muñoz}, \citenamefont {Berga},\ and\ \citenamefont
  {Escriv\'{a}}(2005)}]{Munoz:05}%
  \BibitemOpen
  \bibfield  {author} {\bibinfo {author} {\bibfnamefont {D.~R.}\ \bibnamefont
  {Muñoz}}, \bibinfo {author} {\bibfnamefont {S.~C.}\ \bibnamefont {Berga}}, \
  and\ \bibinfo {author} {\bibfnamefont {C.~R.}\ \bibnamefont {Escriv\'{a}}},\
  }\bibfield  {title} {\enquote {\bibinfo {title} {Current loop generated from
  a generalized impedance converter: A new sensor signal conditioning
  circuit},}\ }\href {\doibase 10.1063/1.1921451} {\bibfield  {journal}
  {\bibinfo  {journal} {Rev. Sci. Instr.}\ }\textbf {\bibinfo {volume} {76}},\
  \bibinfo {pages} {066103} (\bibinfo {year} {2005})}\BibitemShut {NoStop}%
\bibitem [{\citenamefont {Muñoz}\ \emph {et~al.}(2006)\citenamefont {Muñoz},
  \citenamefont {Moreno}, \citenamefont {Berga}, \citenamefont {Montero},
  \citenamefont {Escriv\'{a}},\ and\ \citenamefont {Ant\'{o}n}}]{Munoz:06}%
  \BibitemOpen
  \bibfield  {author} {\bibinfo {author} {\bibfnamefont {D.~R.}\ \bibnamefont
  {Muñoz}}, \bibinfo {author} {\bibfnamefont {J.~S.}\ \bibnamefont {Moreno}},
  \bibinfo {author} {\bibfnamefont {S.~C.}\ \bibnamefont {Berga}}, \bibinfo
  {author} {\bibfnamefont {E.~C.}\ \bibnamefont {Montero}}, \bibinfo {author}
  {\bibfnamefont {C.~R.}\ \bibnamefont {Escriv\'{a}}}, \ and\ \bibinfo {author}
  {\bibfnamefont {A.~E.~N.}\ \bibnamefont {Ant\'{o}n}},\ }\bibfield  {title}
  {\enquote {\bibinfo {title} {Temperature compensation of wheatstone bridge
  magnetoresistive sensors based on generalized impedance converter with input
  reference current},}\ }\href {\doibase 10.1063/1.2358696} {\bibfield
  {journal} {\bibinfo  {journal} {Rev. Sci. Instr.}\ }\textbf {\bibinfo
  {volume} {77}},\ \bibinfo {pages} {105102} (\bibinfo {year}
  {2006})}\BibitemShut {NoStop}%
\bibitem [{\citenamefont {Montero}\ \emph {et~al.}(2007)\citenamefont
  {Montero}, \citenamefont {Muñoz}, \citenamefont {Moreno}, \citenamefont
  {Barrio},\ and\ \citenamefont {Mustelier}}]{Montero:07}%
  \BibitemOpen
  \bibfield  {author} {\bibinfo {author} {\bibfnamefont {E.~C.}\ \bibnamefont
  {Montero}}, \bibinfo {author} {\bibfnamefont {D.~R.}\ \bibnamefont {Muñoz}},
  \bibinfo {author} {\bibfnamefont {J.~S.}\ \bibnamefont {Moreno}}, \bibinfo
  {author} {\bibfnamefont {J.~F.}\ \bibnamefont {Barrio}}, \ and\ \bibinfo
  {author} {\bibfnamefont {A.~S.}\ \bibnamefont {Mustelier}},\ }\bibfield
  {title} {\enquote {\bibinfo {title} {Signal conditioning for differential
  temperature measurement with thermistors using a generalized impedance
  converter},}\ }\href {\doibase 10.1063/1.2778621} {\bibfield  {journal}
  {\bibinfo  {journal} {Rev. Sci. Instr.}\ }\textbf {\bibinfo {volume} {78}},\
  \bibinfo {pages} {086114} (\bibinfo {year} {2007})}\BibitemShut {NoStop}%
\bibitem [{\citenamefont {Borodjieva}\ and\ \citenamefont
  {Manukova-Marinova}(2004)}]{Ruse:04}%
  \BibitemOpen
  \bibfield  {author} {\bibinfo {author} {\bibfnamefont {A.~N.}\ \bibnamefont
  {Borodjieva}}\ and\ \bibinfo {author} {\bibfnamefont {A.~V.}\ \bibnamefont
  {Manukova-Marinova}},\ }\bibfield  {title} {\enquote {\bibinfo {title}
  {Analysis and design of active filters with generalized impedance
  converter},}\ }in\ \href {\doibase 10.1109/ISSE.2004.1490842} {\emph
  {\bibinfo {booktitle} {27th International Spring Seminar on Electronics
  Technology: Meeting the Challenges of Electronics Technology Progress,
  2004.}}},\ Vol.~\bibinfo {volume} {3}\ (\bibinfo  {publisher} {IEEE},\
  \bibinfo {year} {2004})\ pp.\ \bibinfo {pages} {398--404}\BibitemShut
  {NoStop}%
\bibitem [{\citenamefont {Mishonov}\ \emph {et~al.}(2021)\citenamefont
  {Mishonov}, \citenamefont {Petkov}, \citenamefont {Dimitrova}, \citenamefont
  {Serafimov},\ and\ \citenamefont {Varonov}}]{PDF}%
  \BibitemOpen
  \bibfield  {author} {\bibinfo {author} {\bibfnamefont {T.~M.}\ \bibnamefont
  {Mishonov}}, \bibinfo {author} {\bibfnamefont {E.~G.}\ \bibnamefont
  {Petkov}}, \bibinfo {author} {\bibfnamefont {I.~M.}\ \bibnamefont
  {Dimitrova}}, \bibinfo {author} {\bibfnamefont {N.~S.}\ \bibnamefont
  {Serafimov}}, \ and\ \bibinfo {author} {\bibfnamefont {A.~M.}\ \bibnamefont
  {Varonov}},\ }\bibfield  {title} {\enquote {\bibinfo {title} {Probability
  distribution function of crossover frequency of operational amplifiers},}\
  }\href {\doibase https://doi.org/10.1016/j.measurement.2021.109509}
  {\bibfield  {journal} {\bibinfo  {journal} {Measurement}\ }\textbf {\bibinfo
  {volume} {179}},\ \bibinfo {pages} {109509} (\bibinfo {year} {2021})},\
  \Eprint {http://arxiv.org/abs/1802.09342v3} {arXiv:1802.09342v3 [eess.SP]}
  \BibitemShut {NoStop}%
\bibitem [{\citenamefont {Mishonov}\ and\ \citenamefont
  {Varonov}(2021)}]{Varna}%
  \BibitemOpen
  \bibfield  {author} {\bibinfo {author} {\bibfnamefont {T.~M.}\ \bibnamefont
  {Mishonov}}\ and\ \bibinfo {author} {\bibfnamefont {A.~M.}\ \bibnamefont
  {Varonov}},\ }\bibfield  {title} {\enquote {\bibinfo {title} {Scientific
  instrument for creation of effective cooper pair mass spectroscopy},}\ }\href
  {\doibase 10.1088/1742-6596/1762/1/012013} {\bibfield  {journal} {\bibinfo
  {journal} {J. Phys.: Conf. Ser.}\ }\textbf {\bibinfo {volume} {1762}},\
  \bibinfo {pages} {012013} (\bibinfo {year} {2021})},\ \Eprint
  {http://arxiv.org/abs/2009.12315} {arXiv:2009.12315 [cond-mat.supr-con]}
  \BibitemShut {NoStop}%
\end{thebibliography}%

\onecolumngrid

\clearpage
\appendix

\section{Supplementary Experimental Data}

\subsection{Signal + Noise Input and Output from Oscilloscope Output}

\begin{figure*}[h]%
\centering
\includegraphics[scale=4.0]{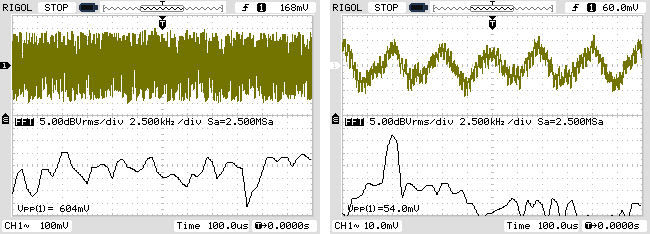}
\caption{Two screenshots from digital oscilloscope (Rigol  DS1052E) screen of a typical application of a resonance filter performed by GIC.
Left: A sum of
white noise (1~V peak to peak) and
small sinusoidal signal (50~mV peak to peak, $f_\mathrm{res}=5.05$~kHz), 
signal to noise ratio 1/20,
is applied from a signal generator (Rigol DG1022) to sequentially connected load resistor (100~k$\Omega$) and GIC,
this is the input signal of the filter.
Right: Voltage on the GIC in this voltage divider, this is the output signal of the filter. 
The small sinusoidal signal is recovered by suppressing the 
noise outside of the resonance.
In technical applications of measurement of small signals
Q-factor of the resonator 
multiplies the dynamic diapason of the lock-in voltmeter.
The output signal passes through a voltage repeater with TL072 operational amplifier 
because of the giant modulus impedance of the GIC in the resonance,
much larger than the input impedance of the oscilloscope (1~M$\Omega$),
see Fig.~3.
Both oscilloscope screens are vertically divided in half:
the upper part represents the time dependence of the voltages,
while the lower part gives the spectral density of the signals
mathematically calculated by the oscilloscope from the voltage signal
shown in the upper part.
On the left one can see approximately constant spectral density of white noise,
the small sine signal cannot be seen even in the spectral density.
On the right the resonance maximum becomes visible due to suppressing of the 
non-resonance frequencies.
The scale of all figures is different
and it can be easily seen that the recovered sinusoidal signal
has the same amplitude as the input sinusoidal signal.
This recovered sine signal dominates the output signal spectral density.
}
\label{Fig:Osc}
\end{figure*}%

\subsection{PCB Layout}

\begin{figure*}[h]%
\centering
\includegraphics[scale=0.32]{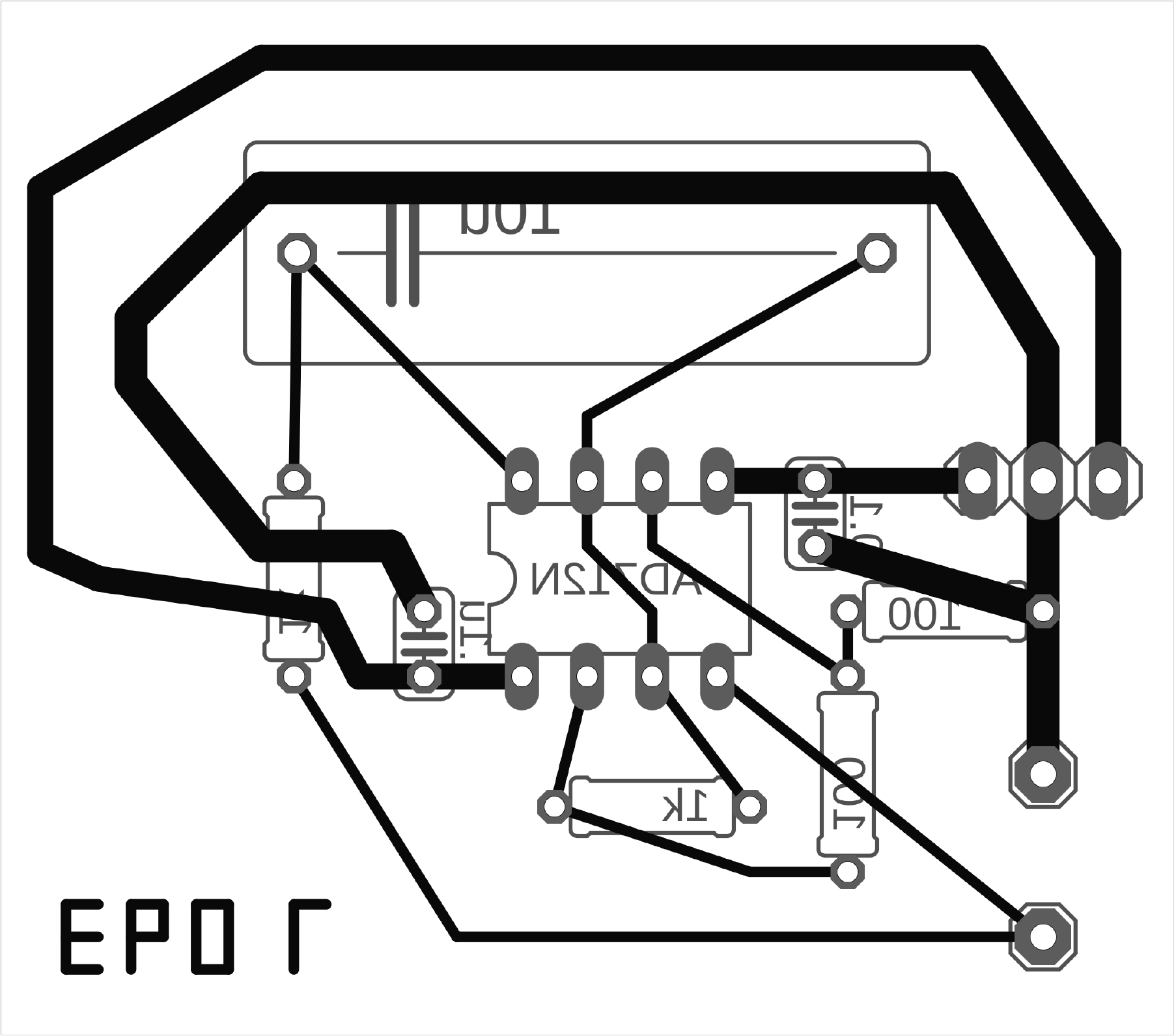}
\caption{Printed Circuit Board (PCB) of the the resonator performed by GIC;
topology is depicted in Fig.~1. 
One can see the place for the big metal-layer capacitor 
$C=10\,\mu\mathrm{F}$; $Z_4=1/\jm\omega C$,
the places of the small resistors $Z_2=Z_1=100\,\Omega$,
and the places of the big resistors $Z_5=Z_3=1\,\mathrm{k}\Omega$.
Not shown in Fig.~1: places for small ceramic capacitors capacitors 
which are connected to the voltage supply batteries are close 
to the 8 pin locus for the operational amplifier.
The 3 pins on the right are for 
voltage supply $V_\mathrm{S-}$,
floating (not connected) common point, and
voltage supply $V_\mathrm{S+}$,
This set-up was given to the participants of the 
7-th Experimental Physics Olympiad, see~Ref.~[12].
At low frequency below 50~Hz high students measured 
that it is an artificial inductance $L=10$~H.
The new idea of the present study is to demonstrate that this 
GIC has inherent high-Q resonance which is perfectly described by
the single pole approximation Eq.~(1)
of the frequency dependent open-loop gain.
The novelty of our result is that this opportunity has never been used to 
create a tunable high-Q resonator.
Our motivation is to create a new set-up for measurement
of small signals in the physics of superconductivity.
The two pins on the lower right part of figure
are the 2 electrodes of the GIC used to connect it in a circuit. 
} 
\label{Fig:PCB}
\end{figure*}%

\section{Poles and Zeros in the Complex Frequency Plane}

\begin{figure*}[h]%
\centering
\includegraphics[scale=0.7]{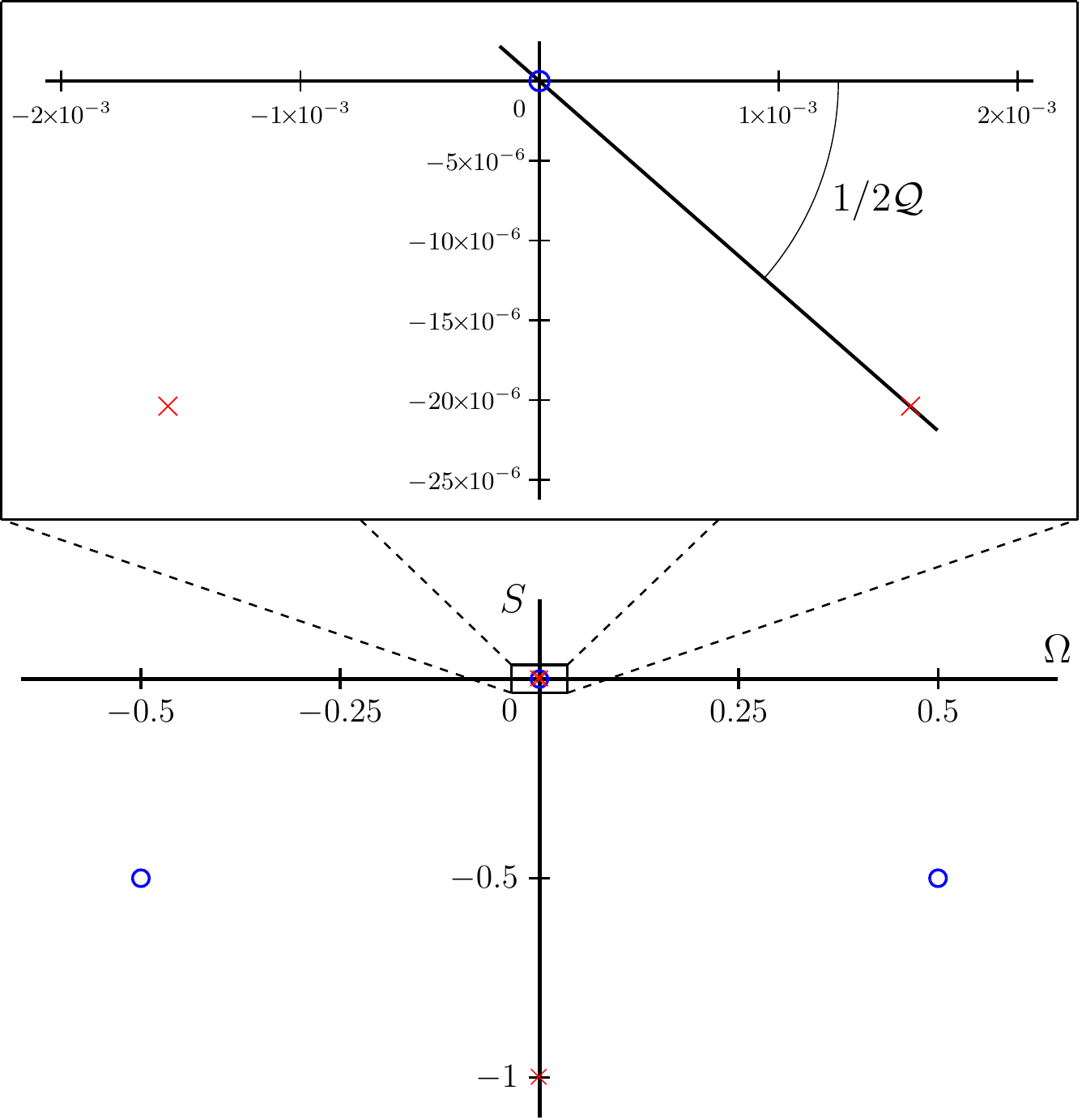}
\caption{Poles (\textcolor{red}{$\star$}) and zeros (\textcolor{blue}{$\circ$}) of the GIC impedance $Z(\omega)$ in the complex plane of the frequency $\Omega\equiv\omega\tau$, where $\omega\equiv -\mathrm{i}s =\omega^\prime+\mathrm{i}\omega^{\prime\prime}$, 
all of them in the lower semi-plane.
The zero $\omega_0\approx -\mathrm{i}R_L/L$ describes current decay $\propto\mathrm{e}^{-(R_L/L)t}$ of DC current through the simulated inductance.
The two pole resonances $\omega_\pm\approx\pm\omega_\mathrm{res}
-\mathrm{i}\omega_\mathrm{res}/2\mathcal{Q}$ describe 
time decay of the voltage amplitude 
$\propto \mathrm{e}^{-(\omega_\mathrm{res}/2\mathcal{Q})t}
\mathrm{e}^{-\mathrm{i}\omega_\mathrm{res}t}$
and energy of oscillations
$\propto \mathrm{e}^{-(\omega_\mathrm{res}/\mathcal{Q})t}$.
The other zeros and pole are irrelevant for the low frequency behavior of the GIC.
For real frequencies and sinusoidal voltages 
$U(t)=\Re(U_\omega
\mathrm{e}^{-\mathrm{i}\mathrm{\omega}t})$ 
and currents
$I(t)=\Re(I_\omega
\mathrm{e}^{-\mathrm{i}\mathrm{\omega}t})$ 
the impedance $Z(\omega) \equiv U_\omega/I_\omega$ 
and the conductivity $\sigma(\omega)\equiv 1/Z(\omega)=I_\omega/U_\omega$ 
describe linear responses of the system with respect to small perturbations.
For damping modes of a stable system $Z(\omega)$ and $1/Z(\omega)$
are analytical functions in the upper $\omega$ semi-plane and one can calculate the Fourier
transforms to time domain 
$Z(t)=\int_{-\infty}^{+\infty}\mathrm{e}^{-\mathrm{i}\omega t}
Z(\omega)(\mathrm{d}\omega/2\pi)=\theta(t)Z(t)$
and
$\sigma(t)=\int_{-\infty}^{+\infty}\mathrm{e}^{-\mathrm{i}\omega t}
\sigma(\omega)(\mathrm{d}\omega/2\pi)=\theta(t)\sigma(t)$.
The Heaviside $\theta$-function describes the causality principle:
$I(t)=\int_0^{\infty}\sigma(t^\prime)U(t-t^\prime)\mathrm{d}t^\prime$
and analogously 
$U(t)=\int_{-\infty}^{t}Z(t-t^\prime)I(t^\prime)\mathrm{d}t^\prime$.
For more details related to Kramers and Kronig causality principles, see for example the
section on generalized susceptibility from the textbook on statistical physics by Landau and Lifshitz.
The amplitude of the plane wave in optics is 
$\propto\exp[\mathrm{i}(\mathbf{k}\cdot\mathbf{r}-\omega t)]$, where the relation $\mathrm{j}=-\mathrm{i}$ comes from.
Let $\zeta>0$ is a positive variable, $s=\zeta$ 
and $\omega=\mathrm{i}\zeta$ is purely imaginary.
In this case 
$Z(\omega=\mathrm{i}\zeta)
=\int_{0}^{+\infty}\mathrm{e}^{-\zeta t}
Z(t)\,\mathrm{d}t.$
If the impedance is a passive system in thermal equilibrium with temperature $T$
the Matsubara frequency is discrete $\zeta_n=2n\pi k_\mathrm{B}T/\hbar$,
where $n=0,1,2,\dots .$
}
\end{figure*}

\section{Alternative Enumeration of the GIC Impedances}

In Fig.~\ref{Fig:GIC} the impedances are numbered as 
floors of a building from the ground upwards. 
However in Fig.~8.45 of Ref.~\onlinecite{Zumbahlen@GIC} and the numbering is opposite, 
as rows of a matrix from up to down, i.e. enumeration (1, 2, 3, 4, 5) from the used above notation should be substituted by (5, 4, 3, 2, 1).
In the enumeration used by Zumbahlen~\cite{Zumbahlen@GIC} our main results
Eq.~(\ref{Z(omega)}) reads
\begin{align}
Z(\omega)=\dfrac{Z_1}{1-\dfrac{Z_3Z_5-Z_2Z_4}{Z_3Z_5+Z_4Z_5
\left[\alpha Z_2+\beta Z_3+\alpha\beta(Z_2+Z_3)\right]}}
\;.
\end{align}

Additionally, here we wish to point out that Fig.~8.46 and Figs.~8.47 A, B and C 
of Ref.~\onlinecite{Zumbahlen@GIC} have erroneous topology,
which is corrected in Fig.~8.48. 

Concerning the terminology in Ref.~\onlinecite{Zumbahlen@GIC} the
notion: ``general impedance converter'' is used,
while some authors recommend ``generalized impedance converter''.
We do not express an opinion but just a comparison 
``general relativity'' or ``generalized relativity'' has to be called the Einstein theory 
for the geometrodymanics. 

\end{document}